\magnification=\magstep1
\tolerance 500
\bigskip
\centerline{\bf Dynamical Relativistic Systems and the Generalized}
\centerline{{\bf Gauge Fields of Manifestly Covariant
Theories}\footnote{*}{Based on a lecture
Presented at the conference on {\it Physical
 Interpretations of Relativity Theory}, British Society for the Philosophy of
Science, London 11-14 September, 1998.}}
\bigskip
\centerline{L.P. Horwitz\footnote{**}{Also at Department of Physics,
Bar Ilan University, Ramat Gan, Israel. E-mail larry@ccsg.tau.ac.il.}}
\centerline{ School of Physics}
\centerline{ Raymond and Beverly Sackler Faculty of Exact Sciences}
\centerline{Tel Aviv University, Ramat Aviv, Israel}
\bigskip
\noindent {\it Abstract\/}: The problem of the classical
non-relativistic electromagnetically
kicked oscillator can be cast into the form of an iterative map on
phase space.  The original work of Zaslovskii {\it et al} showed that
the resulting evolution contains a stochastic flow in phase space to
unbounded energy.  Subsequent studies have formulated the problem in
terms of a relativistically charged particle in interaction with the
electromagnetic field.  We review the standard derivation of the
covariant Lorentz force, and review the structure of the relativistic
 equations used to study this problem.
 We show that the Lorentz force equation
  can be derived as well from
the manifestly covariant mechanics of Stueckelberg in the presence of
a standard Maxwell field.  We show how this agreement is achieved, and
criticize some of the fundamental assumptions underlying these
derivations.  We argue that a more complete theory, involving
``off-shell'' electromagnetic fields should be utilized. We then
discuss the formulation of the off-shell electromagnetism implied by
the full gauge invariance of the Stueckelberg mechanics (based on its
quantized form), and show that a more general class of physical
phenomena can occur.
\bigskip
\noindent
{\bf I. Introduction}
\smallskip
\par In recent years, there has been considerable interest in
classical relativistic dynamical systems, engendered, in part, by the
results of Zaslovskii {\it et al} [1] on the electromagnetically kicked
oscillator.  In this system, a model for real phenomena in plasmas,
it was shown that (under certain conditions) there is an Arnol'd type
diffusion along a stochastic web in phase space, and that the energy
goes to infinity.  The equation studied by Zaslovskii {\it et al}, for
a non-relativistic system in a periodically $\delta$-function kicking
 electric field with transverse magnetic field is
$$ {d^2 x \over dt^2} + \omega_H^2 \,x(t)= {e \over m} E(x,t),
\eqno(1.1)$$
where the electric field was taken to be
$$ E(x,t) = -E_0 T\, sin(k_0x -\omega_0 t) \sum_{n=-\infty}^\infty
\delta(t-nT), \eqno(1.2)$$
and $\omega_H = {eB_0} /mc $ is the Larmor (cyclotron) frequency.  For
rational ratios of the frequency ${2\pi}/T$ and the oscillator
frequency $\omega_H$, the phase space $(x, dx/dt)$ of the system is
covered  by a mesh of finite thickness for which the space within the
cells contains regular dynamics, and the filaments contain motion of a
stochastic nature, even for arbitrarily small fields.  The particles
may then diffuse (in a way analogous to Arnol'd diffusion) arbitrarily
far into the region of high energies.
\par Since the energies of such systems are apparently unbounded, some
authors reformulated the problem using relativistic
kinematics. Longcope and Sudan [2] start from the equations
$$\eqalign{ {dv_x \over dt} + {1\over \gamma} v_x {d\gamma \over dt} +
{\omega_H^2\over \gamma^2} x &= {e \over m\gamma} E(x,t)\cr v_y &=
-(\omega_H/\gamma) x. \cr} \eqno(1.3)$$
They take ${\bf B} = (0,0,B_0)$ and ${\bf E} = (E(x,t),0,0)$, where
$$ E(x,t) = E_0T \sin kx \sum_{n=-\infty}^\infty \delta(t-nT). $$
One can set $z=v_z = 0$, as we shall see below.
\par These equations can be obtained from the covariant form of the
Lorentz force [3-5] (we use the metric $(-,+,+,+)$)
$$ {\ddot x}^\mu = {e \over mc} F^\mu_\nu {\dot x}^\nu, \eqno(1.4)$$
where we take, in this section, ${\dot x}^\mu = dx^\mu/ds$, and
$$ds^2 = dt^2 - {d{\bf x}^2 \over c^2} \eqno(1.5)$$
is the square of the proper time interval on the particle world line.
Using the relation
$$ \bigl({dx^\mu \over ds} \bigr)\bigl({dx_\mu \over ds} \bigr) = -c^2,
\eqno(1.6)$$
 actually assumed in the derivation of $(1.4)$ [4], we see that
$$ \bigl({dt \over ds} \bigr)^2 = 1-{v^2 \over c^2}, \eqno(1.7)$$
from which (taking the positive square root)
$$ {dt \over ds} = {1 \over \sqrt{1 - {v^2 \over c^2}}} =
\gamma. \eqno(1.8)$$
This result is consistent with the Lorentz transformation, but appears
to be stronger.  The transformation laws of special relativity are
valid only in inertial frames.  If the source of $(1.8)$ were an
explicit Lorentz transformation, corresponding to a method which is
sometimes used in developing the consequences of $(1.4)$ ([2], [6],
[7]), the second derivative $d^2t / ds^2$ would clearly not have a
reasonable physical interpretation.  The result $(1.7)$ is, however,
the consequence of an identity (Eq.$(1.6)$)
and hence it appears that it can be differentiated with respect to
$s$ [4]. However, we shall argue in the next section that
 the formula $(1.5)$ relating ``proper time'' to the
interval is highly dynamical, and cannot be understood as a relation
with actual proper time if the particle is
accelerating.\footnote{*}{There is no question that $c^2dt^2-d{\bf
x}^2$, picking out the endpoints of the interval from a particle trajectory,
is a Lorentz invariant.For an accelerating particle, the Lorentz
transformation cannot, however, reach an inertial frame for this
particle, and therefore this invariant does not correspond to the
proper time on the particle.} We continue
here to discuss the consequences of $(1.5)$ to show how the dynamical
map constructed by Longcope and Sudan [2] follows.
\par Longcope and Sudan [2] take, using $(1.7)$,
$${d{\bf x} \over ds } = {dt \over ds} {\bf v} = \gamma {\bf v},
\eqno(1.9)$$
and replacing the second derivatives with respect to proper time by
$\bigl( \gamma {d \over dt}\bigr)^2$, $(1.4)$ implies the differential
equation
$$m {d \over dt}(\gamma {\bf v}) = e ({\bf E} + {{\bf v} \over c}
\times {\bf B}), \eqno(1.10) $$
where we have used
$$ {1 \over \gamma} \bigl({d{\bf x} \over ds}, {dt \over ds} \bigr) =
({\bf v}, 1).$$
The three components of $(1.9)$ are then
$$\eqalign{m {d\over dt} (\gamma v_x) &= eE(x,t) + {v_y \over c} B_0 \cr
m{d\over dt} (\gamma v_y) &= -{e \over c} v_x B_0 \cr
 m {d \over dt} (\gamma v_z) &= 0.\cr} \eqno(1.11)$$
It is therefore consistent to take $v_z=z=0$.  Integrating the second
of $(1.11)$, one then finds $(1.3)$.
\par As Longcope and Sudan [2] show by direct calculation using the map
derived from $(1.3)$ ($E(x,t)=0$ between $\delta$-function kicks; one
obtains a map by integrating between kicks and establishing new
initial conditions after the kick by integrating over the
$\delta$-function), one recovers the stochastic web of Zaslavskii {\it
et el} [1] for small velocities, but for larger velocities, the
distribution in phase space becomes stochastic.
\par Let us now re-examine briefly the structure of the covariant
Lorentz force $(1.4)$.  In addition to the vector equation
$$ {\ddot x}^j = {e \over mc} (F^j_0 {\dot x}^0 + F^j_k {\dot x}^k),
\eqno(1.12)$$
one has the time component
$$ {\ddot x}^0 = {e \over mc} F^0_j {\dot x}^j. \eqno(1.13)$$
Directly changing the independent variable from $s$ to $t$, one obtains
$$\eqalign{{d \over ds} &= {dt \over ds} {d \over dt} \cr
{d^2 \over ds^2} &= {d^2t \over ds^2} {d  \over dt} + ({dt \over ds})^2
{d^2 \over dt^2}. \cr} \eqno(1.14)$$
Hence Eq. $(1.12)$ becomes
$$ {d^2 t \over ds^2}v^j + \gamma^2 {d^2 x^j \over dt^2} = \gamma {e
\over m} (E^j + {1 \over c} F^{jk} v^k).$$
From $(1.13)$, we see that
$$ {d^2 t \over ds^2} = {e \over mc^2} \gamma {\bf E} \cdot{\bf v},
 \eqno(1.15) $$
and hence [5,
6]
$${d^2 {\bf x} \over dt^2 } = {1 \over \gamma} {e \over m} ({\bf E} +
{1 \over c} ({\bf v} \times {\bf B}) - {1 \over c^2} {\bf v} ({\bf v}
\cdot {\bf E})),  \eqno(1.16)$$
or, using the identity [8] ${\bf v} \times ({\bf v} \times {\bf E}) =
v({\bf v} \cdot {\bf E}) - v^2{\bf E}$,
$$ {d^2 {\bf x} \over dt^2} = { 1 \over \gamma} {e \over m} [(1-{v^2
\over c^2}) {\bf E} + {1 \over c} ({\bf v} \times ({\bf B} - {{\bf v}
\over c} \times {\bf E}))], \eqno(1.17)$$
one sees that the effective magnetic field is corrected by a
relativistically induced magnetic field. The repeated
acceleration of an electron by an electric field eventually becomes
ineffective in the direction of motion of the particle.  Its
velocity becomes bounded dynamically by the velocity of light.
 Comparing $(1.16)$ and $(1.10)$, the term $(e / \gamma m c^2){\bf
v}({\bf v} \cdot {\bf E})$ must coincide with $({\bf v} /\gamma)
(d\gamma/ dt)$.
\par Landau [5] remarks that
$$ {dE_{kin} \over dt} = mc^2 {d \gamma \over dt} \eqno(1.18)$$
can be shown to be $e({\bf v} \cdot {\bf E})$.  This follows
directly by noting that
$$ {d \over dt}\bigl( 1 - {v^2 \over c^2} \bigr)^{-{1 \over 2}} =
{\gamma^3\over c^2} {\bf v}\cdot {d{\bf v} \over dt}; $$
the result then follows from $(1.17)$.  It is physically quite
reasonable, since $e({\bf v} \cdot {\bf E}) $ can be interpreted as
the rate at which the field does work on the particle.  However, the
interpretation of $mc^2 \gamma$ as the particle energy is derived from
the Lorentz transformation from a state at rest, and is only valid in
an inertial frame.  It is not clear what the derivative of such an
expression, implying a change of velocity, means.
\par We learn, furthermore, from the form of $(1.13)$ and $(1.15)$,
that the variable $t$, as a function of $s$, undergoes acceleration,
involving more than the Lorentz dilation.  It must therefore be
considered a dynamical variable both in classical and quantum
mechanics, as is done in the framework of Stueckelberg [9,10]
\bigskip
\noindent
{\bf II. The Covariant Stueckelberg Formulation}
\smallskip
\par   In order to examine more clearly the assumptions and structure
of the results outlined above, we formulate the problem in
the manifestly covariant framework of Stueckelberg [9,10],
using the standard Maxwell electromagnetic fields. We remark that
Mendon\c ca and Oliveira e Silva [11] have studied the dynamics
generated by a ``super Hamiltonian'', where the energy $E$ and time $t$
are considered as dynamical variables. This formulation is equivalent to
that of refs. [9] and [10].  We shall consider
later the more general pre-Maxwell fields [12].
\par The Stueckelberg theory constitutes a formulation of relativistic
dynamics in terms of forces, or interactions, in a given Lorentz
frame.  The motions, including acceleration (of any order) are not
associated with the motion of a frame and hence the theory is
applicable directly to the many-body problem (Horwitz and Piron [10]).
 The theory is constructed in a manifestly covariant way, and has the
same form in any Lorentz (or Poincar\'e) frame.  It has a symplectic
(Hamiltonian) structure, and constitutes a direct generalization of the
standard non-relativistic dynamics, but on an 8N dimensional phase
space, including the time $t$ and energy $E$ of each particle.  Given
an invariant Hamiltonian $K(x_1, x_2, \dots, x_N, p_1, p_2, \dots ,
p_N)$, where $x_i \equiv x_i^\mu = (ct_i, {\bf x}_i), \, p_i \equiv
p_i^\mu = ({E_i \over c}, {\bf p}_i)$ are four-vectors, the Hamilton
equations are
$$ {\dot x}_i^\mu = {\partial K \over \partial p_\mu} \qquad {\dot
p}_i^\mu = -{\partial K \over \partial x_\mu} , \eqno(2.1)$$
where the dot now indicates differentiation with respect to an invariant
(universal) parameter $\tau$.  For a single free particle, an
effective choice is
$$ K_0 = {p_\mu p^\mu \over 2M}, \eqno(2.2)$$
where $M$ is a parameter with dimensions of mass.  Then,

$$ {\dot x}_\mu = {p^\mu \over M} \qquad {\dot p}^\mu =
0. \eqno(2.3)$$
From the first of these, we see that
$$ {d{\bf x} \over d\tau} = {{\bf p} \over M},\qquad  c {dt \over
d\tau} = {E \over Mc} \eqno(2.4)$$
and hence
$$ {d{\bf x} \over dt} = c^2 {{\bf p} \over E}, \eqno(2.5)$$
consistent with the definition of velocity in special relativity. This
definition, based on directly observable quantities, does not depend
on whether the particle is accelerating or not. The occurrence of
$p^\mu p_\mu$ in the generator $(2.2)$ of evolution, and the existence
of the dynamical variables $t,E$ as independent of ${\bf x},{\bf p}$,
imply that the theory is in principle not on ``mass shell''.  The
quantity $E^2/c^2 -{\bf p}^2$ is a dynamical variable, and its values
are determined by the equations of motion.  In the case of the free
particle, $p^\mu p_\mu \equiv -m^2c^2$ is a constant of the motion, and
may be chosen arbitrarily in terms of initial conditions.
\par Choosing $m^2 = M^2$ (``on-shell'') so that \footnote{*}{It is
sometimes convenient to choose $m=M$ as the Galilean
(non-relativistic) limit of the variable $m$.  The Galilean group
requires a fixed mass for any representation.  In this sense, we use
the terminology ``on-shell''.}
$$ {E^2 \over c^2} - {\bf p}^2 = M^2 c^2,$$
it follows from Eq. $(2.4)$ that
$$ {dt \over d\tau} = {E\over Mc^2 } = {1\over \sqrt{1-c^2{{\bf p}^2
\over E^2}} } = {1 \over \sqrt{1 -{v^2 \over c^2}}}. \eqno(2.6)$$
Modifying $(2.2)$ to represent a minimal gauge invariant interacting
Hamiltonian [9], we write [9]-[11]
$$K = {(p^\mu -{e \over c} A^\mu)(p_\mu - {e \over c} A_\mu) \over 2M}.
 \eqno(2.7)$$
The equation of motion for $x^\mu$ is
$$ {\dot x}^\mu \equiv {dx^\mu \over d\tau} = {\partial K \over
\partial p_\mu} = {p^\mu -{e \over c} A^\mu \over M} \eqno(2.8)$$
and we see that
$$ {dx^\mu \over d\tau} {dx_\mu \over d\tau}= -c^2 \bigl({ds\over
d\tau} \bigr)^2 =  {(p^\mu -{e \over c} A^\mu)(p_\mu - {e \over c}
A_\mu) \over M}, \eqno(2.9)$$
a quantity which (since it is proportional to the $\tau$-independent
Hamiltonian) is strictly conserved.  In fact, this quantity is the
gauge invariant mass-squared:
$$(p^\mu -{e \over c} A^\mu)(p_\mu - {e \over c}A_\mu)= -m^2c^2,
\eqno(2.10)$$
so that
$$ c^2 \bigl({ds \over d\tau }\bigr)^2 = c^2 \bigl({dt \over d\tau}
\bigr)^2 - \bigl({d{\bf x} \over d\tau} \bigr)^2 = {m^2 c^2 \over M^2}
\eqno(2.11)$$
and
$$\bigl({dt \over d\tau} \bigr)^2 = { m^2/M^2 \over {1 - {v^2 \over c^2}}},
\eqno(2.12)$$
where $m^2$ is a constant of the motion.
\par We now derive the Lorentz force from the Hamilton equation
 (this derivation has also been carried out independently by C. Piron [13])
$$\eqalign{ {dp^\mu \over d\tau} &= - {\partial K \over \partial
x_\mu} = {(p^\nu -{e \over c} A^\nu) \over M } \bigl( {e \over c}
{\partial A_\nu \over \partial x_\mu} \bigr) \cr
&= {e \over c} {dx^\nu \over d\tau} {\partial A_\nu \over \partial
x_\mu} . \cr } \eqno(2.13)$$
Since $p^\mu = M {dx^\mu \over d\tau} + {e\over c} A^\mu$, the
left hand side is ($A^\mu$ is evaluated on the particle world line
$x^\nu(\tau)$)
$$ {dp^\mu \over d\tau } = M {d^2x^\mu \over d\tau^2} + {e \over c}
{\partial A^\mu \over \partial x^\nu} {dx^\nu \over
d\tau},\eqno(2.14)$$
and hence
$$M {d^2 x^\mu \over d\tau^2}= {e \over c} \bigl({\partial A_\nu \over
\partial x_\mu} - {\partial A^\mu \over \partial x^\nu} \bigr) {dx^\nu
\over d\tau},$$
or
$$M {d^2 x^\mu \over d\tau^2} = {e \over c} F^\mu_\nu {dx^\nu \over
d\tau}, \eqno(2.15)$$
where $(\partial^\mu \equiv \partial/\partial x_\mu)$
$$ F^{\mu \nu} = \partial^\mu A^\nu - \partial^\nu
A^\mu. \eqno(2.16)$$
The form of $(2.15)$ is identical to that of $(1.4)$, but the
temporal derivative is not with respect to the variable $s$, the
Minkowski distance along the particle trajectory, but with respect to
the universal evolution parameter $\tau$.
 \par One might argue that these should be equal, or at least
proportional by a constant, since the proper time is equal to the time
which may be read on a clock on the particle in its rest frame.  For
an accelerating particle, however, {\it one cannot transform by a
Lorentz transformation, other
than instantaneously, to the particle rest frame}.  The properties of
an accelerating particle in an instantaneous rest frame are not
equivalent to those in an inertial frame.  As pointed out by Mashoon
[14] \footnote{*}{I thank J. Beckenstein for bringing this work
to my attention.}, the usual assumed equivalence is based on a
postulate of ``locality'' by which the state of the particle is
determined by its position and velocity.  He emphasized that
 this postulate is not, in
general, physically correct.  He cites the example of a
particle in interaction with the radiation field of electromagnetism.
Such a particle radiates if it is accelerating, but the corresponding particle
in a comoving
inertial frame does not (in ref. [5], the radiation formula is
derived by going to an instantaneous rest frame for the particle, and
computing the radiation field of a non-relativistic particle with
acceleration in that frame). Due to this inequivalence, the formula
$(2.15)$ appears to have a more reliable interpretation.  The
parameter of evolution $\tau$ does not require a Lorentz transformation to
achieve its meaning.
\par However, since $m^2$ is absolutely conserved by the
Hamiltonian model $(2.7)$, we have the constant relation
$$ ds = {m\over M} d\tau, \eqno(2.17)$$
assuming the positive root (as we shall also do for the root of
$(2.12)$; we do not wish to discuss the antiparticle solutions here).
Eq. $(2.15)$ can therefore be written
$$ m {d^2x^\mu \over ds^2} = {e \over c} F^\mu_\nu {dx^\nu \over ds},
\eqno(1.4')$$
as in $(1.4)$, with the ``proper time''.
\par This result is somewhat puzzling.  We have obtained the Lorentz
force in terms of ``proper time'', a concept which has somewhat
insecure foundations (without an equivalence principle, as in
Einstein's gravity) in an arbitrary force field.  In fact $ds$ cannot
be the proper time interval for the freely falling system; the system is
accelerating, and this quantity is not accessible through a Lorentz
transformation. Since $\tau$ can be understood as the time one can
 observe on a freely falling ideal clock, the correspondence $(2.17)$ seems
 too strong.
\par One can check with $(2.15)$ (or $(1.4)$) that the conservation
law is trivially maintained by the equations of motion.  Since
$F^{\mu\nu}$ is antisymmetric, multiplying $(1.4')$ by ${\dot x}_\mu$
yields zero, but ${\dot x}^\mu{\ddot x}_\mu = {1 \over 2}
 {d\over ds}({\dot
x}^\mu{\dot x}_\mu) = 0$ is just this conservation law.
\par As we have pointed out, Mashoon's [14] counterexample involves
the phenomenon of radiation.  Very clear discussions of the derivation
of radiation corrections to $(1.4)$ are given by Rohrlich [3] and
Sokolov and Ternov [4], based on the original discussion of Dirac
[15].  In these derivations, the identity $(1.6)$ is used in the
analysis of the series expansion, resulting in the well-known
 ${d\over ds}\ddot x^\mu$
 term.  The result contains the Abraham-Lorentz radiation reaction
terms [16]:
$$ m{\ddot x}^\mu = {e\over c} {\dot x}_\nu F^{\mu \nu} + {2 \over 3}
{r_0 \over c} m ( {d\over ds}{\ddot x}^\mu - {1 \over c^2} {\dot x}^\mu {\ddot
x}_\nu {\ddot x}^\nu ), \eqno(2.18)$$
where $r_0 = e^2/mc^2$, the classical electron radius, and the
dots refer here (to the end of this section), as in $(1.4)$, to
 derivatives with respect to $s$.  It follows
from the identity $(1.6)$ that
$$ {\dot x}_\mu{\ddot x }^\mu = 0,\qquad  {\dot x}_\mu {d\over ds}\ddot x^\mu +
{\ddot x}_\mu{\ddot x}^\mu = 0, \eqno(2.19)$$
and hence $(2.18)$ can be written as
$$ m{\ddot x}^\mu = {e \over c} {\dot x}_\nu F^{\mu\nu} + {2 \over 3}
{r_0 \over c}m {d \over ds}\ddot x^\nu (\delta^\mu_\nu + {1 \over c^2} {\dot
x}^\mu {\dot x}_\nu).  \eqno(2.20)$$
The last factor on the right is a projection orthogonal to ${\dot
x}^\mu$ (if ${\dot x}^\mu {\dot x}_\mu = -c^2$), and $(2.20)$ is
therefore consistent with the conservation of ${\dot x}^\mu {\dot
x}_\mu$.  Sokolov and Ternov [4] state that this conservation law
follows ``automatically'' from $(2.18)$, as from $(1.4')$, but it is
apparently only consistent.
\vfill
\break
\noindent
{\bf III. The Pre-Maxwell Fields}
\smallskip
\par It seems remarkable that electromagnetism appears to be
consistent with the use of accelerating frames under some
circumstances, preserving the relation
$$ dt = {ds \over \sqrt{1 - {v^2 \over c^2} }} \eqno(3.1)$$
One may attribute this to the fact that this relation is not a direct
consequence of the Lorentz transformation (although it is clearly
related to it by the invariance of the definition $c^2ds^2 = c^2dt^2 -
d{\bf x}^2$; recall, however, that this $ds$ is not the proper time for
an accelerating system).  In the more dynamical treatment in terms of
Stueckelberg mechanics, we see this relation emerging due to the
dynamical conservation of $m^2$, equivalent to the constancy of
${dx^\mu \over d\tau} {dx_\mu \over d\tau}$.  And yet, the result is
somewhat disturbing.
\par As pointed out by Sokolov and Ternov [4], there is no Lagrangian
or Hamiltonian starting point for deriving $(2.18)$.  It is a result
of a series of assumptions and decisions dealing with the propagators
of the field.  They understand this obstacle as being associated with
the irreversible nature of, in particular, the radiation term
proportional to ${\ddot x}_\nu {\ddot x}^\nu$.  It is therefore very
difficult to verify conservation laws.
\par As we have seen, although $(2.18)$ is consistent with $(1.6)$,
even this equation (in which this identity is used) does not imply,
without using the relation again, as in $(2.20)$, that $(1.6)$ is
necessarily true.
\par The possibility that the relation $c^2 ds^2 = -dx^\mu dx_\mu$,
for $ds$ the actual proper time of the particle, should acquire a
non-trivial metric (instead of the Euclidean Minkowski form) in
acceleration may be equivalent to the requirement that the particle,
as well as the radiation field, go off mass shell during interaction,
and hence invalidates the argument leading to $(2.12)$ with constant
$m$ (see, for example, ref. [17], where it is shown that an effective potential
in the Stueckelberg evolution function, giving rise to an explicit local
mass shift, can be expressed equivalently in terms
 of a conformal metric tensor).
 Stueckelberg [9], in attempting to illustrate pair production and
annihilation in a classical framework, discovered that the Hamiltonian
$(2.7)$ presents an obstacle; the particle world line cannot turn
continuously to negative $dt/d\tau$ if the proper time does not go
through zero (he added an additional vector field interaction to
complete the illustration).
\par The quantum form of the Stueckelberg theory, in which the
Hamiltonian generates evolution in $\tau$ by means of the
Stueckelberg-Schr\"odinger equation (we take $\hbar = c=1$ in the
following)
$$  i {\partial \psi_\tau \over \partial \tau} = K \psi_\tau
\eqno(3.2)$$
presents a structure suggesting such a useful generalization.  The
invariance of the probability density $\vert \psi_\tau (x)\vert^2$ in
spacetime is preserved by $\psi_\tau(x) \rightarrow e^{ie_0\Lambda}
\psi_\tau(x)$, where $\Lambda$ is a pointwise function of $\tau$ and
$x^\mu$. To compensate for the derivatives of $\Lambda$, one must
introduce gauge compensation fields which we shall call
$a^\alpha(x,\tau)$ (see ref. [12] for discussion and further references),
$$ a^\alpha(x,\tau)= \{a^\mu(x,\tau), a^5(x,\tau) \} . \eqno(3.3)$$
The fifth gauge field is required by the derivative $i\partial_\tau$
(as for the non-relativistic Schr\"odinger equation, where the fourth
gauge field $A^0(x)$ is required by the derivative $i\partial_t$).  The
gauge fields transform as
$$ {a'}^\alpha  = a^\alpha + \partial^\alpha \Lambda. \eqno(3.4)$$
The gauge covariant minimal coupling form of $(3.2)$ is then
$$ i {\partial \over \partial \tau} \psi_\tau(x) =\bigl\{ {1 \over 2M} (p_\mu
-e_0 a_\mu)(p^\mu -e_0a^\mu) - e_0 a_5\bigr\} \psi_\tau(x). \eqno(3.5)$$
It follows from this equation , in a way analogous to the
Schr\"odinger non-relativistic theory, that there is a current
$$ j_\tau^\mu = -{i \over 2M} \{\psi_\tau^*(\partial^\mu -ie_0a^\mu)
\psi_\tau - \psi_\tau(\partial^\mu + ie_0a^\mu)\psi_\tau^* \},
\eqno(3.6)$$
which, with
 $$\rho_\tau \equiv j^5_\tau = \vert \psi_\tau(x)\vert ^2, $$
satisfies
$$ \partial_\tau \rho_\tau + \partial_\mu j_\tau^\mu \equiv
\partial_\alpha j^\alpha = 0. \eqno(3.7)$$
We see that for $\rho_\tau \rightarrow 0$ pointwise ($\int
\rho_\tau(x) d^4x = 1$ for any $\tau$),
$$ J^\mu(x) = \int_{-\infty}^\infty j_\tau^\mu(x) d\tau \eqno(3.8)$$
satisfies
$$ \partial_\mu J^\mu(x) = 0, \eqno(3.9)$$
and can be a source for the standard Maxwell fields.
\par We understand the operator on the right hand side of $(3.5)$ as
the quantum form of a classical evolution function
$$ K = {1 \over 2M} (p_\mu
-e_0 a_\mu)(p^\mu -e_0a^\mu) - e_0 a_5. \eqno(3.10)$$
It follows from the Hamilton equations that
$$ {dx^\mu \over d\tau} = {p^\mu -e_0a^\mu \over M}  \eqno(3.11)$$
and
$$ {dp^\mu \over d\tau} = e_0 {dx^\nu \over d\tau} {\partial a_\nu
\over \partial x_\mu} + e_0 {\partial a_5 \over \partial x_\mu}.$$
Hence
$$ M{d^2 x^\mu \over d\tau^2} = e_0 {dx^\nu \over d\tau} f_\nu^\mu +
e_0 \bigl( {\partial a_5 \over \partial x_\mu} - {\partial a^\mu \over
\partial \tau} \bigr) . \eqno(3.12)$$
If we define $x^5 \equiv \tau$, the last term can be written as
$\partial^\mu a_5 - \partial_5 a^\mu = f^\mu_5$ ,
so that
$$ M {d^2x^\mu \over d\tau^2} = e_0 {dx^\nu \over d\tau}
f_\nu^\mu + e_0 f_5^\mu. \eqno(3.13)$$
Note that in this equation, the last term appears in the place of the
radiation correction terms of $(2.18)$. It plays the role of a generalized
electric field.  Furthermore, we see that the
relation $(1.6)$ consistent with
 the standard Maxwell theory no longer holds as an
identity; the Stueckelberg form of this result $(2.10)$ in the
presence of standard Maxwell fields, where $m^2$ is conserved, is also
not valid.  We now have
$$ {d \over d\tau } {1 \over 2}M \bigl( {dx^\mu \over d\tau}{ dx_\mu
\over d\tau}\bigr) = e_0  {dx^\mu \over d\tau} f_{\mu 5}, \eqno(3.14) $$
with the physical interpretation that the field $f_{\mu 5} $ does work
on the system, resulting in a mass change.   We see that, in general,
 $ {dx^\mu \over d\tau}{ dx_\mu
\over d\tau}$ is not conserved.
\par Writing a Lagrangian that yields, upon variation, the equations
of motion $(3.13)$, and adding the kinetic term
$$ -{\lambda \over 4} \int d\tau\, d^4x \, f^{\alpha \beta} (x,\tau)
f_{\alpha \beta} (x,\tau), \eqno(3.16)$$
one obtains the equations of motion
$$\partial_\beta f^{\alpha \beta} = {e_0 \over \lambda} {dx^\alpha
\over  d\tau} \delta^4(x - x(\tau)) \equiv  e j^\alpha (x,\tau),
\eqno(3.17)$$
where we have identified ${e_0/\lambda}$ with the Maxwell electric charge
$e$ (see below).
  Note that we have raised and lowered the index $\alpha$; the
signature for raising and lowering the fifth index may be $\pm $,
 and we leave this unspecified here.  The current
$j^\alpha$ is the classical analog of the quantum mechanical current
appearing in $(3.7)$.  Considering the $\mu$-components of $(3.17)$,
and integrating over $\tau$ (assuming that $f_\tau^{\mu 5}
\rightarrow 0$ for $\tau \rightarrow\pm \infty$ pointwise), we find
that, with $(3.8)$ (see, e.g., [3] p.81 for the classical case),
$$  \partial_\mu F^{\mu \nu} = e J^\nu, \eqno(3.18)$$
where we have identified
$$ \int d\tau \, a^\mu(x,\tau) = A^\mu(x) \eqno(3.19)$$
with the Maxwell field.  We see that if $A^\mu$ has dimension $L^{-1}$
(reciprocal length), then $a^\mu$ has dimension $L^{-2}$.  The charge
$e_0$ therefore has dimension $L$, as has $\lambda$. It then follows
that ${e_0 / \lambda}$ is dimensionless, and may (classically)
be identified with the Maxwell charge, as indicated in $(3.17)$.
\vfill
\break
\noindent
{\bf IV. Discussion}
\smallskip
\par The greater generality of the pre-Maxwell-Lorentz equations
appears to admit a deeper study of the radiation process; we are presently
applying it to the kicked relativistic oscillator for which the
perturbation may be taken to be $f^{5 \mu}$ (whose source is local
event density in spacetime) or $f^{0j}$ (whose zero-mode is the
Maxwell electric field).  The $\tau$-dependence of the fields, and the
spacetime dependence of $a_5$, makes it possible for the particle to
move off-shell during the radiation process.
\par In conclusion, we remark that in the quantized version of
convential electrodynamics, one thinks of the photon, the quantum of
radiation contained in the field $A^\mu(x)$, as being emitted or
absorbed by an accelerating charge (as in Compton scattering), or
during the change of state of an atom.  The spacial size scales of the
emitter or absorber is of order $10^{-8}$ cm. or smaller, and yet the
photon is a (nonlocalizable if massless) plane wave.  It seems much
more reasonable to think of quantized radiation as being carried by
off-shell (massive) photons [12], which can be very localized during
emission and absorption, and only asymptotically acquire masses in
the neighborhood of zero (in the standard theory, an off-shell photon
is considered as ``virtual''; this property is understood in that the
photon was emitted by some charged particle source and must be
eventually absorbed).  The standard QED teaches us that emission and
absorption vertices are not trivially pointlike [18].  We see that
both Einstein's relativity and quantum electrodynamics suggest a
generalization of Maxwell's electrodynamics and its interaction with
charged particles.  The Stueckelberg formulation of relativistic
dynamics, along with the pre-Maxwell fields that it implies, may
provide a useful model for studying these effects.
\bigskip
\noindent
{\it Acknowledgements\/}: I wish to thank my colleagues at Los Alamos
National Laboratory and the University of Texas at Austin for
discussions and the opportunity to present these ideas at seminars, and
Y. Ashkenazy, J. Beckenstein, C. Piron and F. Rohrlich for discussions.
\bigskip
\noindent
\frenchspacing
{\bf References}
\smallskip
\item{[1]} G.M. Zaslovskii, M. Yu Zakharov, R.Z. Sagdeev, D.A. Usikov
and A.A. Chernikov, Sov. Phys. JETP {\bf 64}, 294 (1986).;
G.M. Zaslovsky, Chaos, {\bf 1}, 1 (1991).
\item{[2]} D.W. Longcope and R.N. Sudan, Phys. Rev Lett. {\bf 59},
1500 (1987).  See also, H. Karamabadi and V. Angelopoulis,
Phys. Rev. Lett. {\bf 62}, 2342 (1989).  Experiments based on a
quantum analysis have been proposed by S.A. Gardinar, J.I. Cirac and
P. Zoller, Phys. Rev. Lett. {\bf 79}, 4790 (1997).
\item{[3]} F. Rohrlich, Classical Charged Particles, Addison Wesley,
Reading,  Mass. (1965).
\item{[4]} A.A. Sokolov and I.M. Ternov, {\it Radiation from
Relativistic Electrons}, Amer. Inst. of
Phys. Translation Series, New York (1986).
\item{[5]} L. Landau and E. Lifshitz, {\it The Classical Theory of
Fields}, Addison Wesley, Cambridge, Mass. (1951).
\item{[6]} M.A. Heald and J.B. Marion, {\it Classical Electromagnetic
Radiation}, Saunders College Pub., Harcourt Brace, New York (1995).
\item{[7]}  J.D. Jackson, {\it Classical Electrodynamics}, John Wiley,
New York(1962).
\item{[8]} I thank F. Rohrlich for a personal communication on this
point.
\item{[9]} E.C.G. Stueckelberg, Helv. Phys. Acta {\bf 14}, 372, 588
(1941); {\bf 15}, 23 (1942).
\item{[10]} L.P. Horwitz and C. Piron, Helv. Phys. Acta {\bf 46}, 316
(1973).
\item{[11]} J.T. Mendon\c ca and L. Oliveira e Silva, Phys. Rev. E {\bf 55},
1217 (1997).
\item{[12]} D. Saad, L.P. Horwitz and R.I. Arshansky, Found. of
Phys. {\bf 19}, 1125 (1989); M.C. Land, N. Shnerb and L.P. Horwitz,
Jour. Math. Phys. {\bf 36}, 3263 (1995); N. Shnerb and L.P. Horwitz,
Phys. Rev A{\bf 48}, 4068 (1993).
\item{[13]} Personal communication.
\item{[14]} B. Mashoon, Proc. VIII Brazilian School of Cosmology and
Gravitation, Editions Frontieres (1994); Phys. Rev. A {\bf 47}, 4498
(1993).
\item{[15]} P.A.M. Dirac, Proc. Roy. Soc. London Ser. A, {\bf 167},
148(1938).
\item{[16]} M. Abraham, {\it Theorie der Elektrizit\"at}, vol. II,
Springer, Leipzig (1905).  See ref.[3] for a discussion of the origin
of these terms.
\item{[17]} R.I. Arshansky, L.P. Horwitz and D. Zerzion, Ann. of Israel
Phys. Soc. {\bf 9}, {\it Developments in General Relativity, Astrophysics
 and Quantum Theory}, ed. F.I. Cooperstock, L.P. Horwitz and J. Rosen,
 A. Hilger, London (1990).
\item{[18]} F. Rohrlich, Amer. Jour. Phys. {\bf 65}, 1051 (1997).
\vfill

\bye
\end